\documentclass[twocolumn,prl,revtex4,amsmath,amssymb,superscriptaddress]{revtex4-1}

\newcommand{\av}[1]{\langle #1 \rangle}

\usepackage[english]{babel}
\usepackage{graphicx}                       
\usepackage{dcolumn}                        
\usepackage[colorlinks=false]{hyperref}     
\usepackage{bm}                             
\usepackage{subfigure}
\usepackage{longtable}
\usepackage{mathtools}

\usepackage{amssymb}
\usepackage{amsmath}

\usepackage{color}

\begin{document}
\author{Enrico Ubaldi}
\affiliation{\small Institute for Scientific Interchange Foundation, 10126 Torino, Italy}
\affiliation{\small Dipartimento di Fisica e Scienza della Terra, Universit\`a di Parma, Parco Area delle Scienze 7/A, 43124 Parma, Italy}
\affiliation{\small INFN, Gruppo Collegato di Parma, Parco Area delle Scienze 7/A, 43124 Parma, Italy}

\author{Alessandro Vezzani}
\affiliation{\small Dipartimento di Fisica e Scienza della Terra, Universit\`a di Parma, Parco Area delle Scienze 7/A, 43124 Parma, Italy}
\affiliation{\small Centro S3, CNR-Istituto di Nanoscienze, Via Campi 213A, 41125 Modena Italy}

\author{M{\'a}rton Karsai}
\affiliation{\small Univ de Lyon, ENS de Lyon, INRIA, CNRS, UMR 5668, IXXI, 69364 Lyon, France}

\author{Nicola Perra}
\affiliation{\small Centre for Business Network Analysis, University of Greenwich, Park Row, London SE10 9LS, United Kingdom}


\author{Raffaella Burioni}
\affiliation{\small Dipartimento di Fisica e Scienza della Terra, Universit\`a di Parma, Parco Area delle Scienze 7/A, 43124 Parma, Italy}
\affiliation{\small INFN, Gruppo Collegato di Parma, Parco Area delle Scienze 7/A, 43124 Parma, Italy}

\title{Burstiness and tie reinforcement in time varying social networks}

\date{\today}

\begin{abstract}
We introduce a time-varying network model accounting for burstiness and tie reinforcement observed
in social networks. The analytical solution indicates a non-trivial phase diagram determined by the
competition of the leading terms of the two processes. We test our results against numerical
simulations, and compare the analytical predictions with an empirical dataset finding good
agreements between them. The presented framework can be used to classify the dynamical features of
real social networks and to gather new insights about the effects of social dynamics on ongoing
spreading processes.
\end{abstract}
\maketitle{}

The recent availability of longitudinal and time-resolved datasets capturing social behaviours has induced a paradigm shift in the way we study, describe, and model the interactions
between individuals. It moved the focus from static, time-aggregated, representations to
time-varying, dynamical, characterisations of social
networks~\cite{butts2008relational,Holme:2012lr,holme2015modern,gonccalves2015social}. Thinking in
terms of time-varying systems allows to overcome the limitations
arising from the depiction of social ties as fixed and immutable in
time~\cite{Holme:2012lr,holme2015modern}. Indeed, it allows to capture a set of complex dynamics
driving the evolution of links
~\cite{PhysRevLett.112.118702,isella2011s,grindrod2011communicability,praprotnik2015spectral, ghoshal2006attractiveness,
saramaki2015seconds,saramaki2014persistence} and to uncover the effects of such dynamics on
processes unfolding on the networks' fabrics \cite{PhysRevE.87.032805,PhysRevE.83.045102,PhysRevE.89.032807, valdano2015analytical,
    rocha2013bursts, scholtes2014causality, han2015epidemic,morris1997concurrent, rocha2014random, kivela2012multiscale} (see
Ref.~\cite{holme2015modern} for a recent review).

While social networks are shaped by a multitude of processes~\cite{jackson2008social}, two
particular mechanisms have been found to play central roles in their emergence and
evolution~\cite{Onnela01052007, Karsai:2014aa, PhysRevE.83.045102,
Miritello:2013aa,10.1371/journal.pone.0022656, PhysRevE.89.032807, ubaldi2015asymptotic}. The first is social activity,
i.e. the propensity of individuals to be engaged in social act per unit time. Observations in a
range of real datasets, capturing different types of social dynamics, show that activity is
heterogeneously distributed among
people~\cite{Perra:2012uq,ubaldi2015asymptotic,Vincenzo-Tomasello:2014aa,Ribeiro:2013aa}.
Furthermore, the activity of single individuals evolves through \emph{bursty} temporal patterns,
inducing heterogeneous inter-event time distributions~\cite{Malmgren25112008, PhysRevLett.114.108701, 0295-5075-81-4-48002, PhysRevE.73.036127,
Barabasi:2005aa,Karsai:2012aa, 1367-2630-14-1-013055, 10.1371/journal.pone.0040612,
PhysRevE.83.025102}. In other words, not only individuals show heterogeneous propensities to be
socially active, but their activation is bursty as well.

The second important mechanism determines the \textit{allocation of social ties}, i.e. the selection
process driving the creation or renewal of a particular connection. Intuitively, social tie
allocation is not random. In fact, empirical observations show that people tend to distribute a large
fraction of their social acts towards already existing strong ties, while allocating a smaller
amount of interactions to create new social relationships or to maintain weak
ties~\cite{Granovetter1983-GRATSO-3,Onnela01052007,Karsai:2014aa,ubaldi2015asymptotic,Miritello:2013aa,weng2015attention}.
In other words, in time some connections are frequently reinforced by repeated interactions, while
others are not. Although the study of these mechanisms has been the focus of a range of
works~\cite{Onnela01052007,Karsai:2014aa,ubaldi2015asymptotic,PhysRevE.83.045102,
Miritello:2013aa,10.1371/journal.pone.0022656,PhysRevE.89.032807}, a general modeling framework is
still missing. Such a framework would allow for the analytical characterization on how the interplay
of heterogeneous activity patterns and tie allocation mechanisms shape the evolution of social
networks, and in turn the processes taking place on their fabrics.

Here we introduce a model of time-varying networks that allows to regulate the relative strength of
burstiness and tie reinforcement simultaneously. We analytically solve the asymptotic behaviour of
the model and find a non-trivial phase diagram determined by the interplay of the two processes
under investigation.
In particular, we observe two different dynamical regimes, one in which burstiness governs the
evolution of the network, and vice versa another in which the dynamics is completely determined by
the process of ties allocation. Interestingly, if the reinforcement of previously activated
connections is sufficiently strong, burstiness can be sub-leading the network evolution even in the
presence of large inter-event time fluctuations. The theoretical results are validated considering
a real world social network, with analytical predictions fitting the empirical observations.
Consequently, the proposed framework can be also used to classify the temporal features of real
networks, which could provide new insights on the effects that social mechanisms have on spreading
processes unfolding on social networks.

\begin{figure}
    \centering
    \includegraphics[width=.75\columnwidth]{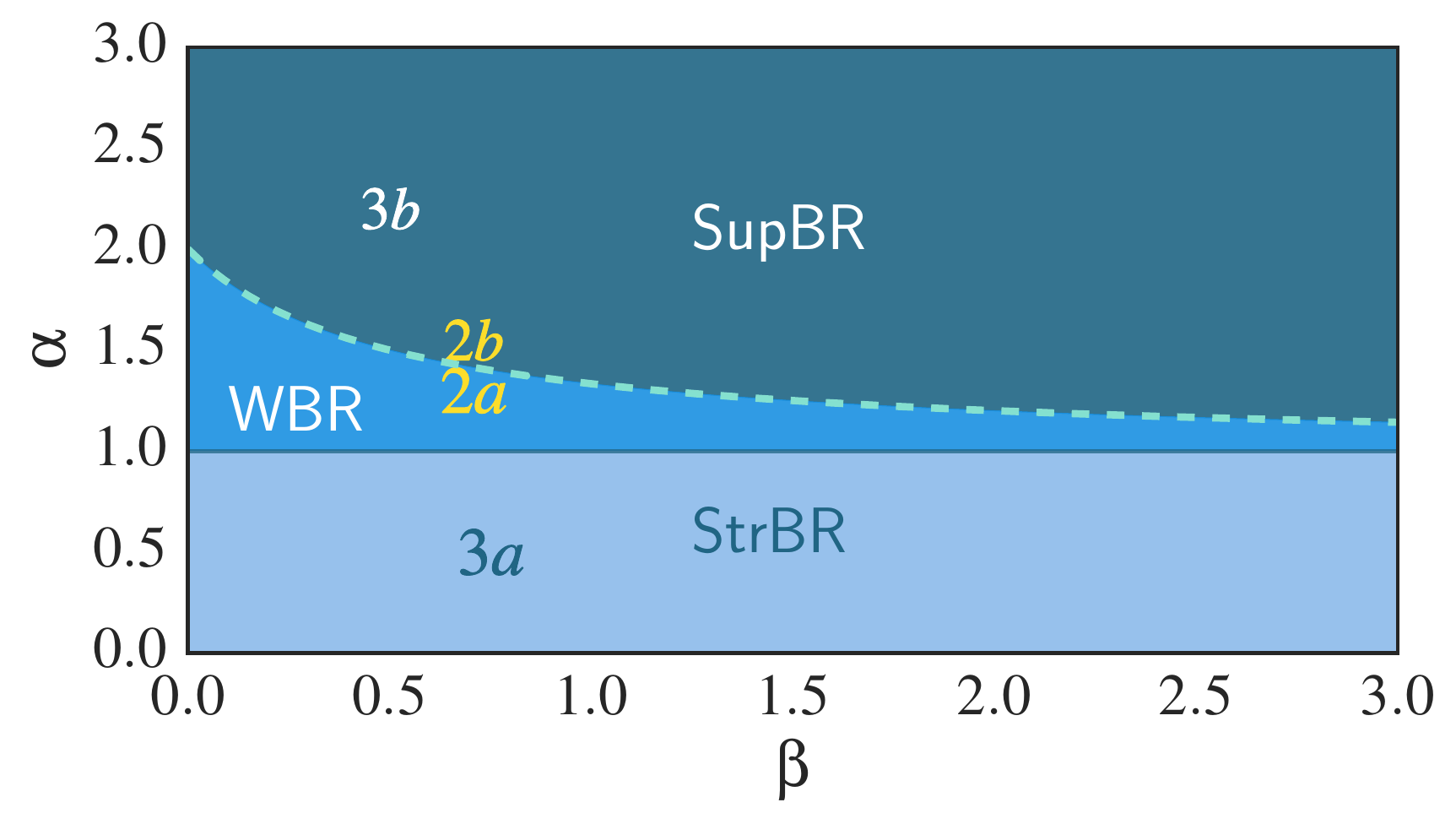}
    \caption{
        \label{fig:bounds} In the phase diagram we report the delimiting lines of the different
        scaling regions as found in Eq. (\ref{eq:Pkt_total}). Evidencing the \textbf{Str}ong
        \textbf{B}urstiness \textbf{R}egime (StrBR), the \textbf{W}eak \textbf{B}urstiness
        \textbf{R}egime (WBR), and the \textbf{Sup}pressed \textbf{B}urstiness \textbf{R}egime
        (SupBR). We also show the parameters value of the simulations presented in Fig.
        \ref{fig:scaling_threshold} (yellow tags), and in Fig. \ref{fig:pakt_sim} (white and blue
        tags).
    }
\end{figure}

In the framework of \emph{activity driven
networks}~\cite{Perra:2012uq,Karsai:2014aa,ubaldi2015asymptotic}, a network $\mathcal{G}$ is
formed by $N$ nodes and each node $i$ is assigned with an activity $a_i$ drawn from an arbitrary
distribution $F(a)$. The activity sets the activation rate of node $i$, i.e. the probability
$a_i\Delta t$ that a node active in time interval $\Delta t$. In general, $F(a)$ is chosen to be a
broad function reflecting the shape of the corresponding distribution in empirical
observations~\cite{Perra:2012uq,Karsai:2014aa,ubaldi2015asymptotic}. Typically 
a power law distribution i.e. $F(a)\sim a^{-(\nu+1)}$ is observed for large activity.

The inter-event time $\tau_i$, i.e. the time between two subsequent activations of the agent $i$,
is directly connected with the agent activity, since $a_i=1/\langle \tau_i \rangle$.
All activity-driven models proposed so far considered a Poissonian distributed $\tau_i$  
~\footnote{With exception for \cite{PhysRevLett.114.108701} which however does not
	consider the reinforcement process}.  However, in social systems the inter-event time
distribution of a single agent is strongly heterogeneous and usually spans over several orders of
magnitude. In order to capture this bursty nature of human dynamics, we impose that the
inter-event time $\tau_i$ for node $i$ is drawn from a power-law distribution $\Psi(\tau_i)$:
\begin{equation}
	\Psi(\tau_i) = {\alpha\over \xi_i^{-\alpha}}\tau_i^{-(1 + \alpha)}, \;\; \tau_i \in
	[\xi_i, +\infty),
	\label{eq:bursty}
\end{equation}
where the exponent $\alpha$ characterizes the distribution and $\xi_i$ is a lower time cutoff. The
latter represents the characteristic timescale for the node $i$, i.e. $\xi_i\sim 1/a_i$, as the
$\gamma$-th moment of the distribution $\Psi(\tau_i)$ reads $\av{\tau_i^\gamma} \sim \xi_i^\gamma$.
If also the $\xi_i$ are heterogeneously distributed as
\begin{equation}
	\Phi(\xi_i) \propto \xi_i^{\nu-1},
	\label{eq:tmin_dist}
\end{equation}
for small $\xi_i$, as a consequence we obtain a network in which the corresponding activity
potential $a_i$ is broadly distributed. In particular the activity distribution behaves as $F(a_i)
\propto a_i^{-(\nu+1)}$ for large $a_i$. We note that, instead of introducing an agent dependent cut-off,  
different definitions are possible, e.g. considering a distribution
of waiting times $\Psi(\tau_i)=\alpha \delta_i^\alpha/(\delta_i+\tau_i)^{1+\alpha}$, where
$a_i \sim 1/\delta_i$  since  $\av{\tau_i^\gamma} \sim \delta_i^\gamma$.
Our model, therefore, belongs to a novel class of activity driven networks, where the agent time 
scale is set by a parameter in the waiting time distribution.

\begin{figure}
    \centering
        {\includegraphics[width=.85\columnwidth]{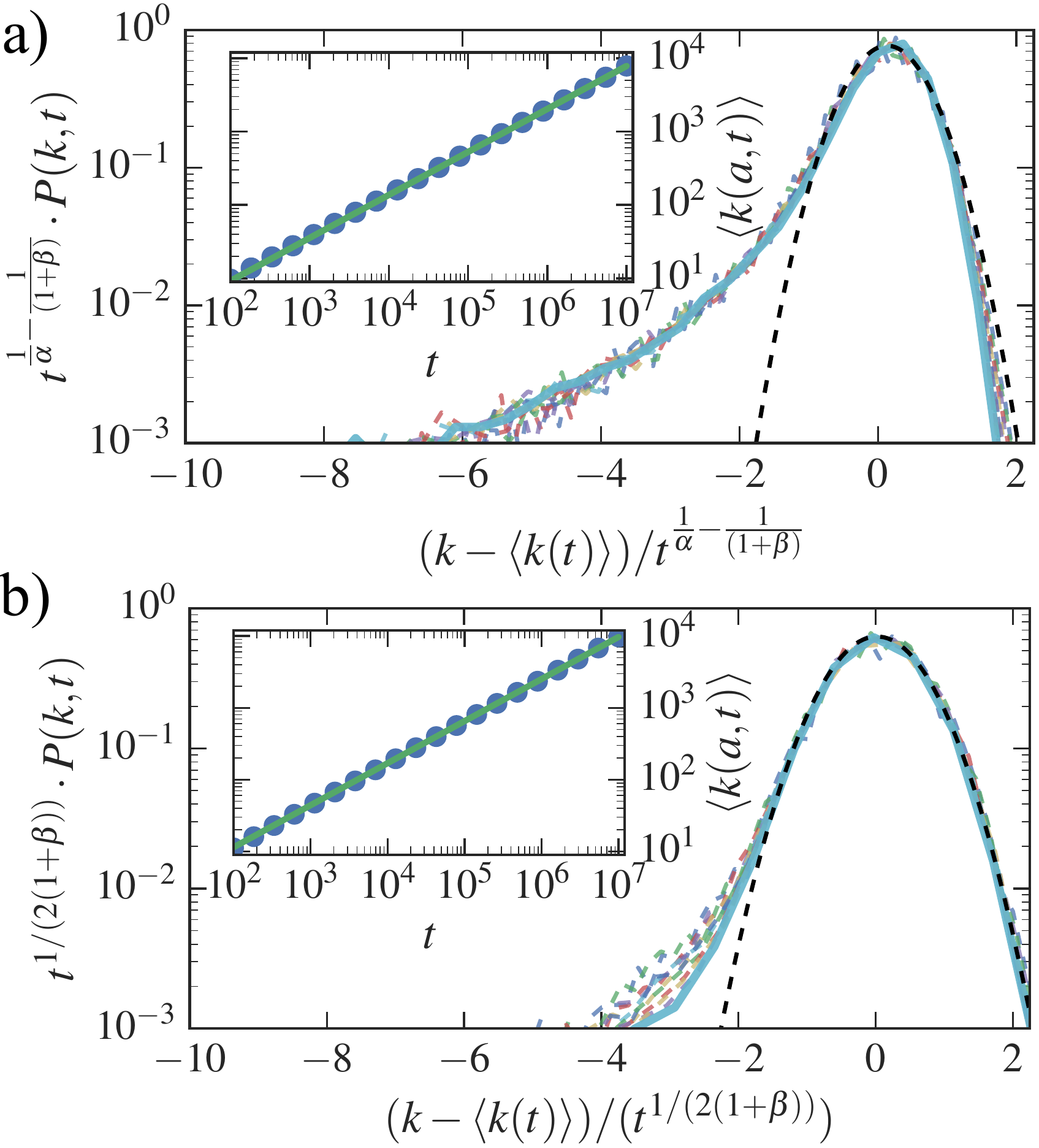}}
    \caption{
        \label{fig:scaling_threshold} (color online) The scaling of the $P(k,t)$ function for (a)
        $\beta=0.7,\,\alpha=1.35$ (WBR region) and (b) $\beta=0.7,\,\alpha=1.6$ (SupBR regime). In
        each plot we consider logarithmically spaced times between $t=10^4$ and $t=10^7$ averaged
        over $10^5$ representation of the single agent evolution. The curves referring to the
        longest time $t=10^7$ are shown in solid thick line, while shorter times are shown in dashed
        lines. A comparison with a normal distribution (black dashed lines) evidence a good
        agreement with the SupBR data (b) while it completely misses the WBR case (a). Insets plot
        the $\av{k(t)}$ and the corresponding analytical prediction of Eq. (\ref{eq:kt_total})
        (green solid lines).
    }
\end{figure}

When a node is active, it initiates interactions with other nodes in the network. This way the degree
$k_i$ of a node $i$, defined as the number of connected peers of $i$ up to time $t$, is evolving in
time. To model this evolution we build on the latest development of the model in which the selection
of ties is driven by a reinforcement process~\cite{Karsai:2014aa,ubaldi2015asymptotic}. In
particular, if at time $t$ a node $i$ of degree $k_i$ is active it will contact a new, randomly
chosen node with probability $p_i(k_i)$. Instead, with probability $1-p_i(k_i)$ it reinforces a tie
by contacting a node randomly chosen amongst the $k_i$ already contacted agents. The form of
$p_i(k_i)$ has been measured and characterized~\cite{ubaldi2015asymptotic} in several datasets as:
\begin{equation}
    p_i(k_i) = \left( 1 + \frac{k_i}{c_i} \right)^{-\beta_i},
    \label{eq:pn}
\end{equation}
where $\beta_i$ measures the strength of the reinforcement process, and $c_i$ sets the
characteristic number of ties that an individual is able to maintain before the reinforcement
process takes place. Though simple, the reinforcement mechanism provides a reliable description of
real world datasets and allows for analytical tractability.

In our simulations, initially, for each node $i$ we set the integrated degree
$k_i=0$ and assign a lower cut-off $\xi_i$ according to distribution $\Phi(\xi_i)$ in  Eq.
(\ref{eq:tmin_dist}). The evolution starts by extracting, for each node $i$, a time $\tau_i$ at
which the node will get active for the first time.  We then activate one node at a time accordingly
to their next activation time. When active, an agent $i$ selects a randomly chosen other agent in
the network with probability $p(k_i) = (1+k_i/c)^{-\beta}$; in this case the value of $k_i$ is
increased by one both for the connecting and the connected nodes.  Otherwise, with probability
$1-p_i(k_i)$, the agent $i$ interacts with a randomly chosen neighbor node which has been already
connected to $i$.  For simplicity we fix $\beta$ and $c$ constant for all nodes. After each
iteration the interaction of node $i$ is removed and a new activation time is selected by an
inter-event time $\tau_i$ drawn from the distribution in Eq. (\ref{eq:bursty}).

\begin{figure} \centering {\includegraphics[width=.85\columnwidth]{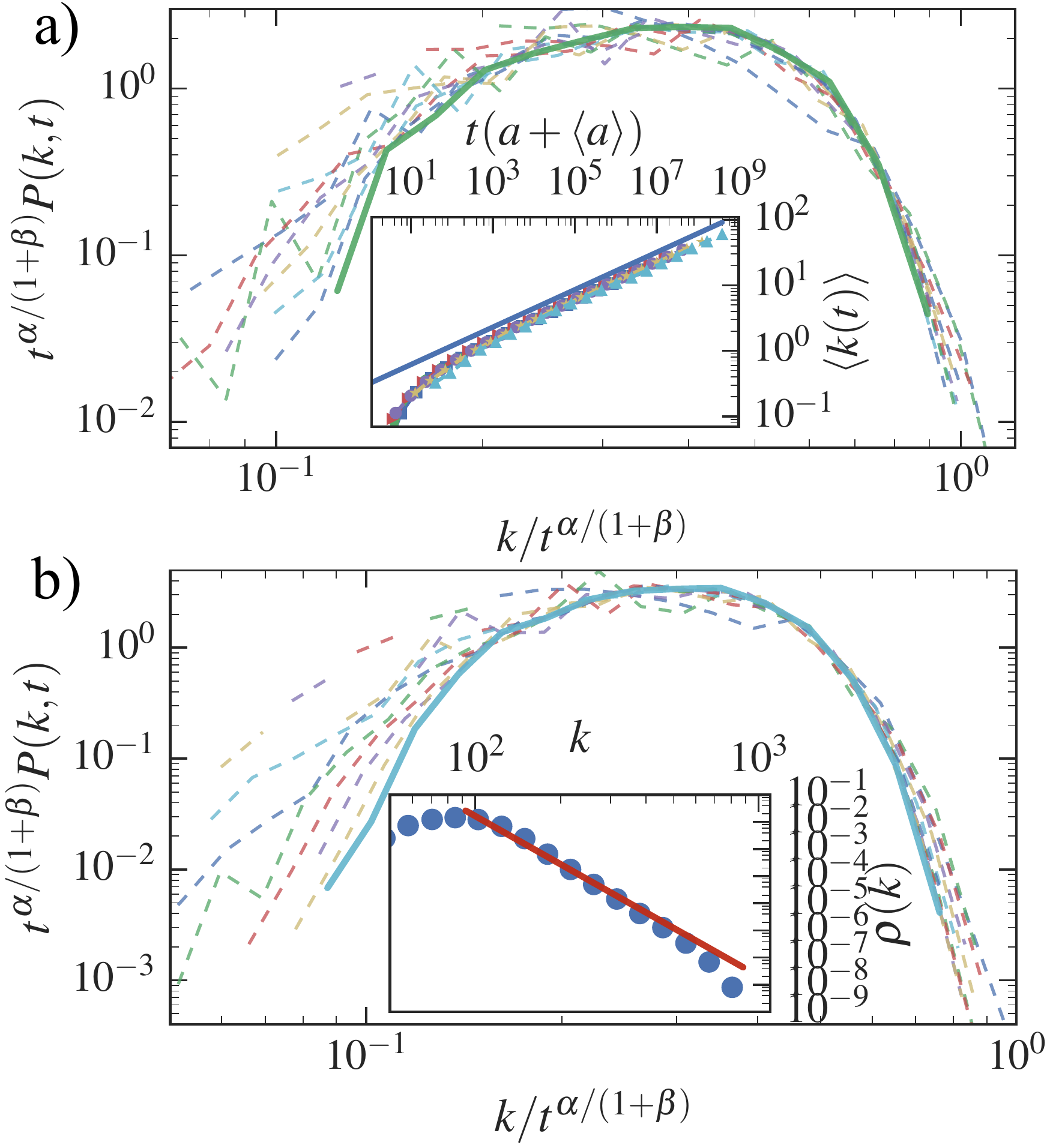}} \caption{
        \label{fig:pakt_sim} (color online). Numerical simulation of the full multi agent dynamics.
        (a) the rescaled $P(k,t)$ distribution for $\xi_i$ distributed with $\nu = 1.4$,
        $\alpha=0.5$, $\beta=0.75$. Curves refer to ten logarithmically spaced times between
        $t=10^5$ and $t=10^8$ (dashed lines, the longest time is shown in solid line). The data
        correspond to the StrBR regime. In the inset we show the growth of average degree for
        different activity classes (symbols) rescaled as $t\to t(a+\av{a})$. The analytical
        prediction of Eq. (\ref{eq:kt_total}) is shown for comparison (blue solid line). (b) the
        $P(k,t)$ distribution for $\nu=1.2$, $\beta = 0.5$ and $\alpha=2.2$, at seven
        logarithmically spaced times between $10^4\le t\le 10^6$.  Inset compares the degree
        distribution at the final simulation time (circles) with the analytical prediction (solid
        line) of Eq.(\ref{eq:rho_k}).
    }
\end{figure}

In the following we apply a single agent approximation, in which agents can only attach to other
nodes and never get contacted. In this case we can write the master equation (ME) describing their
degree evolution, and analytically solve it in the asymptotic limit of large times.
In this case the activity is fixed to the value $a_0=\xi_0^{-1}$ where $\xi_0$ is the
characteristic time of the considered agent.
In particular, let us define $Q(k,t)$ as the probability for an agent active at time $t$ to have
degree $k$ right after $t$. The ME then reads as
\begin{equation}
    \begin{split}
    Q(k,t) =  \int_{\xi_0}^\infty \frac{Q(k-1,t-t')}{t'^{\alpha+1}}
        \frac{c^\beta}{(c+k-1)^\beta} +\\
    \int_{\xi_0}^\infty \frac{Q(k,t-t')}{t'^{\alpha+1}}
    \left(1- \frac{c^\beta}{(c+k)^\beta}\right)+ \delta(k,0)\delta(t,0).
    \end{split}
    \label{eq:dyn}
\end{equation}
The first term accounts for the probability that the agent gets active and contacts a new node, while
the second term represents the probability of connecting to an already contacted neighbor. We
evaluate the probability  $P(k,t)$ for a node to have degree $k$ at the time $t$, by integrating Eq.
(\ref{eq:dyn}) over all the possible waiting time values, i.e.
\begin{equation}
    P(k,t) = \int_{\xi_0}^{\infty}
        {Q(k,t-t')\int_{t'}^{\infty}{1\over\tau^\alpha} dt d\tau}.
    \label{eq:Pakt_ME}
\end{equation}

In the asymptotic regime $P(k,t)$ is (see Supporting Materials for details):
\begin{widetext}
\vspace{-.2in}
\begin{equation}
    P(k,t)\simeq
        \begin{cases}
            {1 \over (t a_0)^{\alpha\over1+\beta}}
            f_{\alpha\beta}\left( A'_{\alpha,\beta}
            {k \over (t a_0)^{\alpha\over1+\beta}} \right)
            \,&\rm{if}\,\alpha<1,\\
            {1\over (t a_0)^{ {1\over \alpha}-{1\over(1+\beta)} }}
            f_{\alpha\beta}\left( A'_{\alpha,\beta}{k-v(t a_0)^{1/(1+\beta)}
            \over (t a_0)^{ {1\over \alpha}-{1\over(1+\beta)} }}\right) &
            \,\rm{if}\,1<\alpha<{2\beta+2\over2\beta+1},\\
            {1\over (t a_0)^{1\over 2(1+\beta)}}
            \exp{\left[-A_{\beta}\frac{\left(k-C_\beta (t a_0)^{{1\over
            1+\beta}}\right)^2}{(t a_0)^{1/(1+\beta)}}\right]}&
            \,\rm{if}\,\alpha>{2\beta+2\over2\beta+1},\\
        \end{cases}
    \label{eq:Pkt_total}
\end{equation}
\end{widetext}
where $f_{\alpha\beta}(x)$ is a non-Gaussian scaling function (see
\cite{1742-5468-2013-09-P09022}), $v$ is the drift velocity of the peak of the
distribution and $A_{\alpha,\beta}$, $A_\beta$, $C_\beta$ are constants
depending on $\alpha$ and $\beta$, respectively.

As a consequence of Eq. (\ref{eq:Pkt_total}), the growth of the average degree
$\av{k(t)}$ is:
\begin{equation}
    \av{k(t)}\propto
    \begin{cases}
        t^{\alpha/(1+\beta)}&\;\; \rm{if}\; \alpha<1, \\
        t^{1/(1+\beta)}&\;\; \rm{if}\; \alpha>1. \\
    \end{cases}
    \label{eq:kt_total}
\end{equation}

This solution leads to a dynamical phase diagram of the model, summarized in Fig.~\ref{fig:bounds}.
For $\alpha<1$, in the \textbf{Str}ong \textbf{B}urstiness \textbf{R}egime (StrBR) burstiness
strongly affects the dynamics. Here the scaling function $f_{\alpha\beta}(x)$ is not Gaussian and
the exponent $\alpha/(1+\beta)$, leading the growth of $\av{k(t)}$, depends both on the burstiness
and on the reinforcement exponents, $\beta$ and $\alpha$ respectively. On the other hand, for
$\alpha > (2\beta+2)/(2\beta+1)$, we have a \textbf{Sup}pressed \textbf{B}urstiness \textbf{R}egime
(SupBR), where the dynamics is independent of $\alpha$. In particular, the reinforcement driven
behavior, described in reference \cite{ubaldi2015asymptotic}, is fully recovered with a Gaussian
scaling function and a connectivity growing as $t^{1/(1+\beta)}$. Finally, for
$1<\alpha<(2\beta+2)/(2\beta+1)$ the average connectivity grows as $t^{1/(1+\beta)}$ just as in the
systems with suppressed burstiness. In this regime, the scaling function is not Gaussian and its scaling
length depends on the burstiness exponent $\alpha$. We call this phase the \textbf{W}eak
\textbf{B}urstiness \textbf{R}egime (WBR). The non trivial dependence on $\beta$ and $\alpha$  of
the transition line between WBR and SupBR highlights the complex interplay between burstiness and
tie reinforcement occurring for $1<\alpha<2$. Fig. \ref{fig:scaling_threshold} shows that the curve
$\alpha=(2\beta+2)/(2\beta+1)$ marks a transition from a Gaussian to a non Gaussian scaling
function, providing a numerical support to the analytical asymptotic results. In particular, in Fig.
\ref{fig:scaling_threshold} (a) panel the left tail of the curve is slowly increasing with time, and
the asymmetric distribution cannot be fitted with a Gaussian. On the other hand, in Fig.
\ref{fig:scaling_threshold} (b) we observe the opposite behavior: the long time curve is slowly
converging to a normal PDF.

\begin{figure}
    \centering
        {\includegraphics[width=.85\columnwidth]{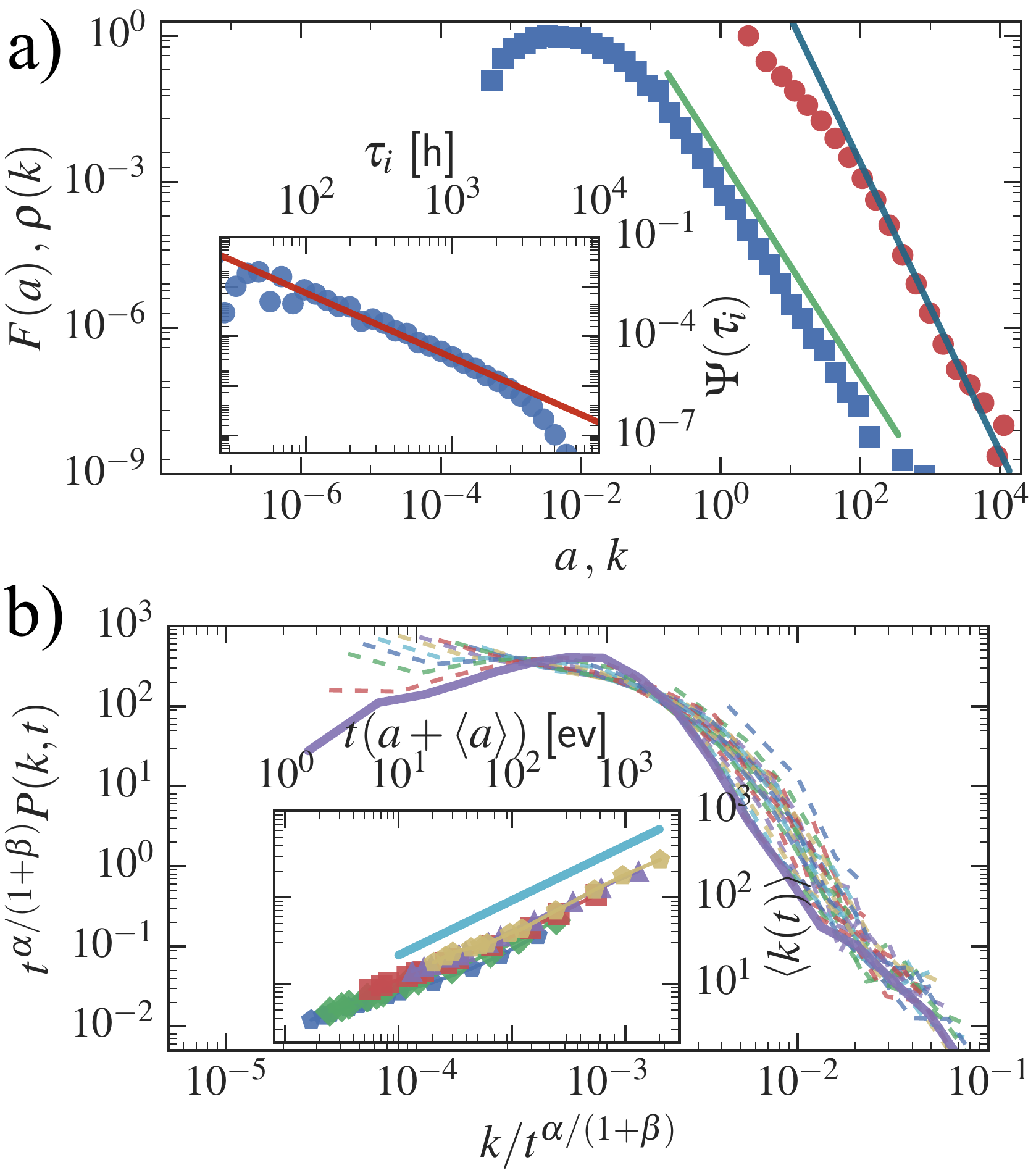}}
    \caption{
        \label{fig:Pakt_data} (color online). Twitter mentions dataset. (a) The activity
        distribution $F(a)$ (blue squares) fitted as $F(a)\propto a^{-(1+\nu)}$ (green solid line)
        with $\nu=1.25$ and the degree distribution $\rho(k)$ (red circles) with the predicted
        behavior (blue solid line) of Eq. (\ref{eq:rho_k}). In the inset we plot the waiting-time
        distribution $\Psi(\tau_i)$ (blue circles) and the fit $\Psi(\tau_i)\propto
        \tau_i^{-(1+\alpha)}$ (red solid line) giving $\alpha = 0.95$. (b) The distribution
        $P(a,k,t)$ for a selected activity class, the degree $k$ is rescaled dividing by
        $t^{\alpha/(1+\beta)}$ where $\beta = 0.47$ has been found in \cite{ubaldi2015asymptotic}
        and $\alpha$ was evaluated in the upper panel. Inset shows the average degree growth
        $\av{k(t)}$ for different activity classes $a$ (symbols) rescaling time as $t\to
        t(a+\av{a})$. The predicted behavior of Eq. (\ref{eq:kt_total}) is shown for comparison
        (green solid line).
    }
\end{figure}

Interestingly, the single-agent model provides a qualitatively correct description even of the
multi-agent case, where different agents display different activities (see Fig. \ref{fig:pakt_sim}).
In particular, one can suppose that the degree distribution Eq. (\ref{eq:Pkt_total}) holds for each
node of the system if one replaces $a_0 t$ with $(a+\langle a \rangle)t$. In this case  $a t$ and
$\langle a \rangle t$ represent the contribution to the growth of the degree induced by the node
activity and by the activity of the rest of the network respectively (see
~\cite{ubaldi2015asymptotic} and ~\cite{PhysRevE.87.062807} for an explicit calculation in the case
without burstiness ). Simulations evidence that the approximation works quite well, however, larger
evolution times are required for observing the asymptotic scaling behavior. The same hypothesis
allows to evaluate the degree distribution among different agents.
In particular, if the activity $a$ is distributed according to Eq. (\ref{eq:tmin_dist}), at a given time
$t$ the degrees, for large $k$, are distributed as:
\begin{equation}
    \rho(k)\propto
    \begin{cases}
        k^{-[\nu(1+\beta)/\alpha+1]}&\;\; \rm{if}\; \alpha<1, \\
         k^{-[\nu(1+\beta)+1]}&\;\; \rm{if}\; \alpha>1. \\
    \end{cases}
    \label{eq:rho_k}
\end{equation}
As shown in Fig. \ref{fig:pakt_sim}(b) inset, the simulation results are well described by the
asymptotic behavior in Eq.~(\ref{eq:rho_k}).

To motivate our model with real world observations, we checked how the proposed scaling picture
corresponds to rescaled empirical $P(k,t)$ functions measured in a Twitter Mentions Network (TMN)
datasets (for details see SM). First of all, the system appears with broad activity and degree
distributions (see Fig.~\ref{fig:Pakt_data}(a)) with exponents satisfying the relations obtained in
Eq.~\ref{eq:rho_k}. Moreover, the inter-event time distribution approximately follows a power-law
(see Fig. \ref{fig:Pakt_data}(a) inset). Notice that, given the measured value of the exponent
$\alpha\sim0.95$, we expect the TMN system to fall in the StrBR region. This is verified in Fig.
\ref{fig:Pakt_data}(b) where indeed the $P(k,t)$ distributions at different evolution times are not
Gaussian and seem to apply to the expected dynamical scaling.
The proposed framework allows to classify the dynamical features of real social networks and thus
anticipate their effects on spreading processes taking place on their fabrics.

In summary, we introduced a new model, which is able to capture two key aspects driving the
evolution of social networks: burstiness and tie reinforcement. We solved the ME of the model in the
large time regime and explored analytically a complex phase space, where changes in the relative
importance between the two mechanisms are linked to different degree distributions and emerging
dynamics. Interestingly, the presented framework is able to predict observables of real networks and
provides a new way to classify their dynamical features. Starting from the analytic result,
an interesting further improvement is the introduction
	of a dynamic on the network population, encoding the fact that in real world dataset agents typically can enter or exit from
	the system during the network evolution.
	
	Our results also have the potential to provide new insights for the characterization of spreading
processes unfolding on social networks. In particular, the proposed modeling framework could help to
clarify the role of burstiness on contagion phenomena, which is currently subject of a heated
debate. The model can potentially be extended further by including other social processes such as
the presence of communities or ageing of nodes.


%

\end{document}


\title{Supplemental Material for ``Burstiness and tie reinforcement in time varying social networks''.}

\author{Enrico Ubaldi}
\affiliation{\small Institute for Scientific Interchange Foundation, 10126 Torino, Italy}
\affiliation{\small Dipartimento di Fisica e Scienza della Terra, Universit\`a di Parma, Parco Area delle Scienze 7/A, 43124 Parma, Italy}
\affiliation{\small INFN, Gruppo Collegato di Parma, Parco Area delle Scienze 7/A, 43124 Parma, Italy}

\author{Alessandro Vezzani}
\affiliation{\small Dipartimento di Fisica e Scienza della Terra, Universit\`a di Parma, Parco Area delle Scienze 7/A, 43124 Parma, Italy}
\affiliation{\small Centro S3, CNR-Istituto di Nanoscienze, Via Campi 213A, 41125 Modena Italy}

\author{M{\'a}rton Karsai}
\affiliation{\small Univ de Lyon, ENS de Lyon, INRIA, CNRS, UMR 5668, IXXI, 69364 Lyon, France}

\author{Nicola Perra}
\affiliation{\small Centre for Business Network Analysis, University of Greenwich, Park Row, London SE10 9LS, United Kingdom}


\author{Raffaella Burioni}
\affiliation{\small Dipartimento di Fisica e Scienza della Terra, Universit\`a di Parma, Parco Area delle Scienze 7/A, 43124 Parma, Italy}
\affiliation{\small INFN, Gruppo Collegato di Parma, Parco Area delle Scienze 7/A, 43124 Parma, Italy}

\maketitle

\section{Dataset} 
\label{sec:dataset}

The analyzed dataset is the Twitter fire-hose (i.e. all the citations
done from all the users) from January the 1st to September the 31st of 2008.

The dataset consists of $536,210$ nodes performing about $160$M events and
connected by $2.6$M edges.

Since the data are daily aggregated we infer the inter-event time distribution
for $\tau_i\lesssim 24$h by assuming the events done by a node within a single day
to be homogeneously distributed during the 24 hours of the day.
As we are measuring the $\alpha$ exponent leading the $\Psi(\tau_i)\propto
\tau_i^{-(1+\alpha)}$ in the right tail of the distribution this assumption does
not change the resulting $\alpha$.

To measure the reinforcement process and specifically the $\beta=0.47$ exponent
we measure the attachment probability on nodes featuring similar stories, i.e.
with a comparable activity $a_i$ (i.e. the number of events actually engaged by
the node $i$) and final degree $k_i$ (see \cite{ubaldi2015asymptotic} for details).

We also checked that sorting the nodes accordingly to their average inter-event
time $\xi_i$ and final degree $k_i$ instead of by their activity does not
change the measured value of $\beta$ that reads $\beta=0.50$ instead of
$\beta=0.47$.

\begin{figure}[]
    \centering
    \subfigure[]
        {\includegraphics[width=2.8in]{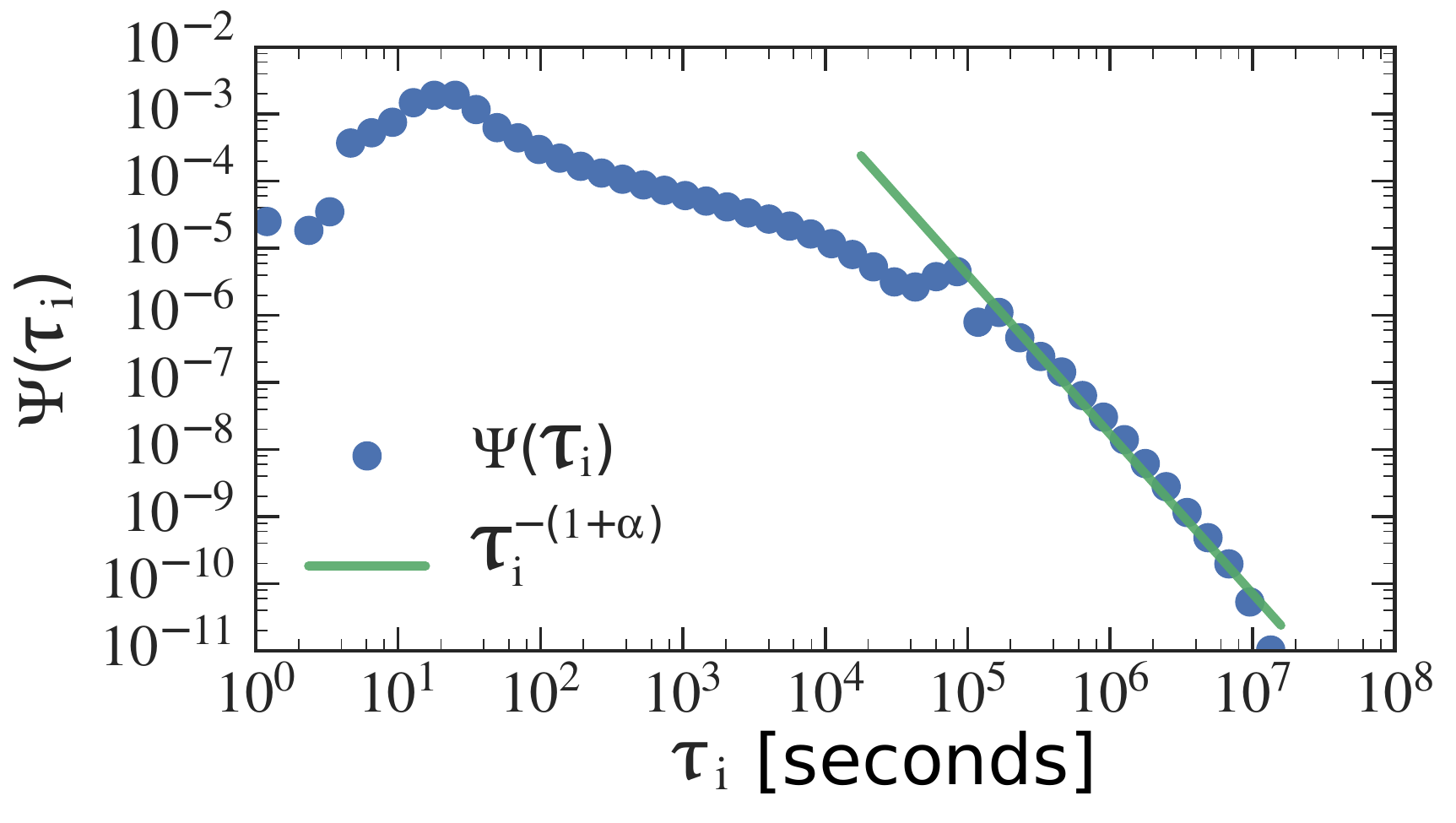}}
    \subfigure[]
        {\includegraphics[width=2.8in]{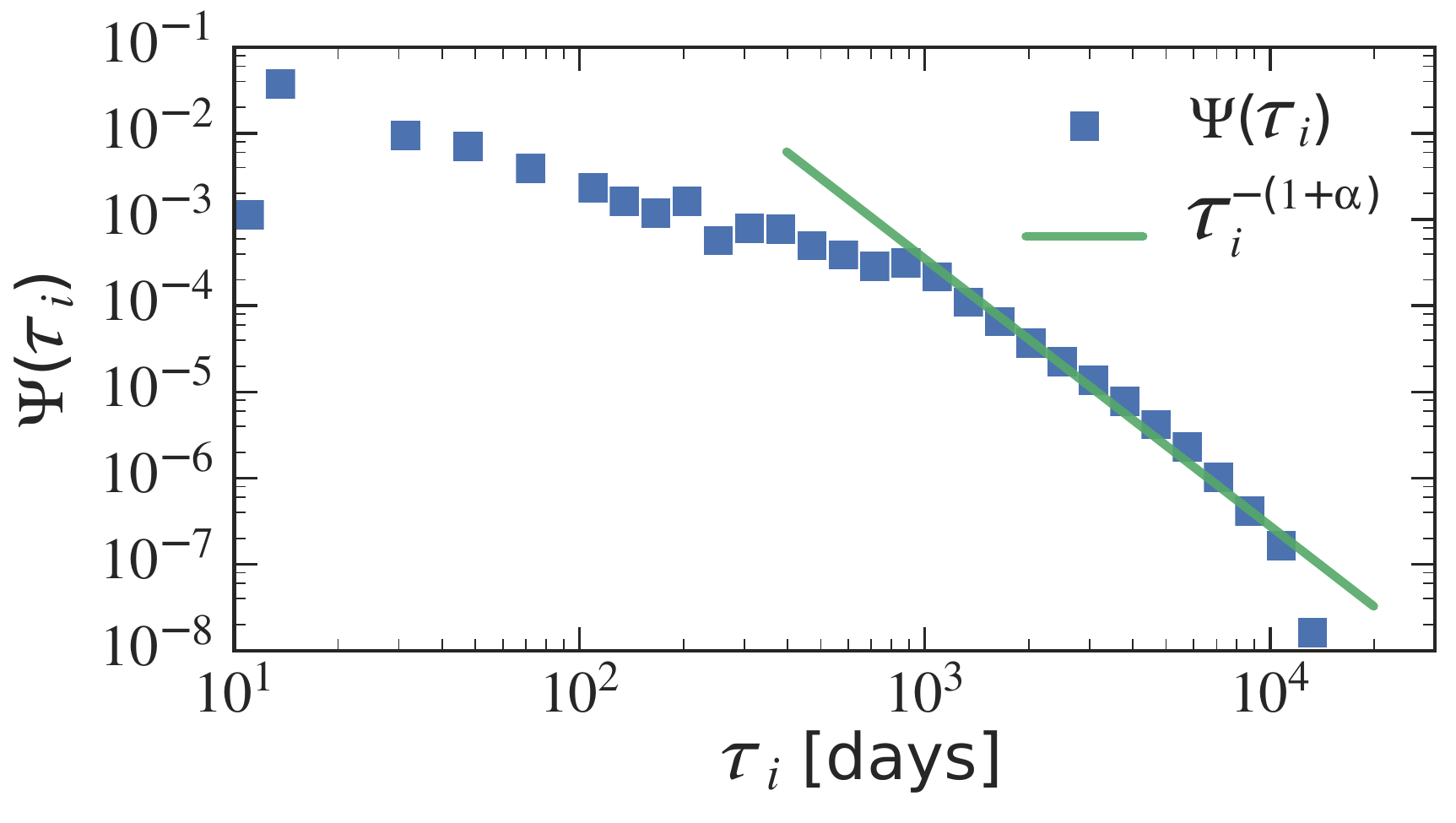}}
    \caption{
        \label{fig:inter_mpc_prb} The waiting-time distribution $\Psi(\tau_i)$  for
        (a) Mobile phone network (circles) and (b) the co-authorship network of
        the Physical Review B journal (squares). We also show the fitting curve
        of the right tail $\Psi(\tau_i)\propto \tau_i^{-(1+\alpha)}$ giving $\alpha
        \sim 1.45$ for the mobile calls dataset and $\alpha \sim 2.1$ for the
        PRB one. Given these results, and provided that the (minimum) value of
        $\beta$ found in the mobile phone call dataset is $\beta_{\rm min}=1.2$
        we conclude that both the system are above the $\alpha =
        (2\beta+2)/(2\beta+1)$ curve, thus falling in the SupBR regime.
    }
\end{figure}

We also considered two other datasets at our disposal (Mobile Phone Calls
network and the American Physical Society co-authorship network) but they both
fall in the SupBR region as we show in Fig. \ref{fig:inter_mpc_prb}.
Given this result we did not further analyzed these datasets as they are already
well described by the previous version of the model.

\section{Master Equation} 
\label{sec:Master Equation}
\subsection{$P(k,t)$} 
\label{sub:$P(k,t)$}

We start from the probability $Q(k,t)$ that an agent makes a call at time $t$
and afterwards its connectivity is $k$.
In the single agent approximation we have:
\begin{equation}
    Q(k,t) =
    \NC \int_{\xi_0}^{+\infty}{ {Q(k-1, t-t') \over t'^{\alpha+1}}
    {c^\beta \over (c+k-1)^\beta} dt'} +
    \NC \int_{\xi_0}^{+\infty}{ {Q(k, t-t') \over t'^{\alpha+1}}
    \left( 1 - {c^\beta \over (c+k)^\beta} \right) dt'} +
    \delta(k,0)\delta(t,0),
    \label{eq:Qkt_1}
\end{equation}
where $\alpha$ is the exponent driving the inter-event time distribution
$P(t) = \NC t^{-(\alpha+1)}$ and $\NC$ is the normalization
constant $\NC = \alpha\xi_0^\alpha$.

To obtain the probability distribution $P(k,t)$ that the agent has degree $k$ at
time $t$ we must integrate Eq. (\ref{eq:Qkt_1}) so that:
\begin{equation}
    P(k,t) = \int_{\xi_0}^{t}{dt' Q(k,t-t') \int_{t'}^{+\infty}
    {d\xi { 1\over\xi^{\alpha+1}}}}.
    \label{eq:Pkt_1}
\end{equation}

Let us perform the Fourier Transform of Eq. (\ref{eq:Qkt_1}) in time, by sending
the integration variable $t\to(t-t')$ we get:
\begin{equation}
    \tilde Q(k,\omega) = \NC \left[
        {c^\beta \over (c+k-1)^\beta} \tilde Q(k-1,\omega)
        \int_{\xi_0}^{\infty}{ {e^{i\omega t'} \over t'^{\alpha+1}} dt'} +
        \left(1 - {c^\beta \over (c+k)^\beta} \right) \tilde Q(k,\omega)
        \int_{\xi_0}^{\infty}{ {e^{i\omega t'} \over t'^{\alpha+1}} dt'} +
    \right] + \delta(k,0).
    \label{eq:Qkw_1}
\end{equation}
By taking the limit $k\to\infty$ of Eq. (\ref{eq:Qkw_1}) we end up with
\begin{equation}
    \tilde Q(k,\omega) = \NC \left[
        \left({c\over k}\right)^\beta \left[ \tilde Q(k-1,\omega)-\tilde Q(k,\omega) \right]
        \int_{\xi_0}^{\infty}{ {e^{i\omega t'} \over t'^{\alpha+1}} dt'} +
        \tilde Q(k,\omega)
        \int_{\xi_0}^{\infty}{ {e^{i\omega t'} \over t'^{\alpha+1}} dt'} +
    \right] + \delta(k,0).
    \label{eq:Qkw_1}
\end{equation}


\subsection{Integral Contributions} 
\label{sub:Integral Contributions}

The issue is now to compute the integral appearing in Eq. (\ref{eq:Qkw_1}).
There are three intervals of the exponent $\alpha$ leading to three different
results. In all the three cases we will perform the integral by taking the
$\omega\to0$ limit, i.e. take into account only the long-time, asymptotic region
of the solution.
\begin{itemize}
    \item $0<\alpha<1$:
        in this regime we can take the $\omega\to0$ limit and expand the
        exponential term $e^{i\omega t} \sim 1 + i\omega t + \mathcal{O}(t^2)$.
        However, the first term proportional to $t$ diverges as $\alpha < 1$. We
        can use the following trick to estimate the first diverging contribution
        of the integral:
        \begin{equation}
            \NC {\partial \over \partial{\omega}} \int_{\xi_0}^{\infty}
            { {e^{i\omega t'} \over t'^{\alpha+1}} dt' } =
            i\NC \int_{\xi_0}^{\infty} { {e^{i\omega t'} \over t'^{\alpha}} dt' },
            \label{eq:intt_al1}
        \end{equation}
        and by defining $|\omega|t' = x$ we get
        \begin{equation}
            \begin{split}
                i\NC
                \int_{|\omega|\xi_0}^{\infty} { {e^{ix \sign(\omega) } \over
                \left({x\over|\omega|}\right)^{\alpha}} {dt' \over |\omega|}} =
                {i\NC |\omega|^{\alpha-1}}
                \int_{|\omega|\xi_0}^{\infty} { {e^{ix \sign(w) } \over
                {x}^{\alpha}} dt'} =\\
                i\NC |\omega|^{\alpha-1}
                \int_{|\omega|\xi_0}^{\infty} { {\cos(x \sign(\omega)) + i \sin(x \sign(\omega))} \over
            x^\alpha} dt' \sim\\
                {i\NC |\omega|^{\alpha-1}} \left[
                {(\xi_0|\omega|)^{1-\alpha} \over \alpha - 1} +
                \Gamma(1-\alpha)\left[ \cos\left( -{\pi\alpha\over2} \right) +
                i \sign(\omega)\sin\left( -{\pi\alpha\over2} \right)\right]
                \right].
            \end{split}
            \label{eq:intt_al2}
        \end{equation}
        Now we can integrate the last term of Eq. (\ref{eq:intt_al2}) in
        $d\omega$ to get the leading term of the integral in the $\omega\to0$
        limit
        \begin{equation}
            \NC \int_{\xi_0}^{\infty}{{e^{i\omega t'} \over t'^{\alpha+1}}dt'}
            \sim
            1 -(\xi_0\omega)^\alpha {\alpha \over \alpha} \Gamma(1-\alpha)
            \left[ \cos\left( -{\pi\alpha\over2} \right) +
            i \sign(\omega)\sin\left( -{\pi\alpha\over2} \right) \right] = 1-
            (\xi_o\omega)^\alpha A_\alpha,
            \label{eq:intt_al3}
        \end{equation}
        where the $1$ comes from the first term of the exponential expansion and
        where we dropped the terms $\propto\omega$ as they die out faster than
        the ones $\propto\omega^\alpha$. We also encoded the complex constant
        multiplying the $(\xi_0\omega)^\alpha$ term in the $A_\alpha$ symbol.
    \item $1<\alpha<2$: in this case we can apply the same procedure of the
        previous case but expanding the series to the second order, thus
        deriving twice with respect to $\omega$.
        The leading terms of the integral of  Eq. (\ref{eq:Qkw_1}) are:
        \begin{equation}
            \NC\int_{\xi_0}^{\infty}{{e^{i\omega t'} \over t'^{\alpha+1}}dt'}
            \sim
            1 + i(\omega\xi_0) {\alpha \over (\alpha-1)}+
            {(\xi_0\omega)^\alpha \over (\alpha-1)}
            \Gamma(2-\alpha) \left[ \cos\left({\pi\alpha\over2}\right) - i
            \sign(\omega) \sin\left({\pi\alpha\over2}\right) \right],
            \label{eq:intt_1a2}
        \end{equation}
        and we define $C_\alpha = {1\over(\alpha-1)} \Gamma(2-\alpha)
        \left[ \cos\left({\pi\alpha\over2}\right) - i \sign(\omega)
        \sin\left({\pi\alpha\over2}\right) \right]$.
    \item $\alpha>2$: in this case it is sufficient to consider the first three
        orders of the integral expansion so that:
        \begin{equation}
            \NC\int_{\xi_0}^{\infty}{{e^{i\omega t'} \over t'^{\alpha+1}}dt'}
            \sim
            1 + i(\omega\xi_0) {\alpha \over (\alpha-1)}-
            (\omega\xi_0)^2 {\alpha \over (\alpha-2)}.
            \label{eq:intt_agt2}
        \end{equation}
\end{itemize}


\subsection{Asymptotic solution $\alpha < 1$} 
\label{sub:Asymptotic solution_al1}

Using Eq. (\ref{eq:intt_al3}) we can rewrite Eq. (\ref{eq:Qkw_1}) as
\begin{equation}
    A_\alpha (\xi_0\omega)^\alpha Q(k,w) +
    \left({c\over k}\right)^\beta
    {\partial Q(k,\omega)\over\partial k} = \delta(k,0).
    \label{eq:Qkw_alt1}
\end{equation}
To solve the equation we introduce the variable $h=k^{1+\beta}$ so that:
\begin{equation}
    A_\alpha (\xi_0|\omega|)^\alpha Q(h,\omega) +
    c^\beta (1+\beta) {\partial Q(h,\omega)\over\partial h} = \delta(h,0).
    \label{eq:Qkw_alt1_1}
\end{equation}
We Fourier transform this equation in space sending $h\to q$ getting
\begin{equation}
    A_\alpha (\xi_0|\omega|)^\alpha Q(q,\omega) +
    i c^\beta (1+\beta) q Q(q,\omega) = 1,
    \label{eq:Qkw_alt1_2}
\end{equation}
so that:
\begin{equation}
    Q(q,\omega) = {1 \over
        A_\alpha (\xi_0|\omega|)^\alpha + i q c^\beta (1+\beta)}.
    \label{eq:Qkw_alt1_3}
\end{equation}

Now let us introduce the variable $h$ also in Eq. (\ref{eq:Pkt_1}) calling
$R(h,\omega) = P(h^{1/(1+\beta)}, \omega)$.
We then perform the FT of Eq. (\ref{eq:Pkt_1}) with respect to $h$ calling
$q$ the transformed variable and getting:
\begin{equation}
    R(q,\omega) = \tilde Q(q,\omega)
    \int{e^{i\omega t'} \int_{t'}^{\infty}{\NC\over \xi^{\alpha+1}} dt'd\xi}.
    \label{eq:Qkw_alt1_1}
\end{equation}
The integral on the r.h.s. of the last equation is the same as the one in Eq.
(\ref{eq:intt_al3}) with $\alpha\to\alpha-1$ so that:
\begin{equation}
    R(q,\omega) = \frac{B_\alpha |\omega|^{\alpha-1}}
    {A_\alpha (\xi_0|\omega|)^\alpha + ic^\beta(1+\beta)q},
    \label{Pkt_alt1_1}
\end{equation}
where $B_\alpha = A_{\alpha-1}\xi_0^\alpha$.
This equation is the same of Eqq. (8) and subsequent in reference
\cite{1742-5468-2013-09-P09022} therefore we have
\begin{equation}
    P(h,t) =\frac 1 {c^\beta (\beta+1)(t/\xi_0)^{\alpha}} f_\alpha\left(
    \frac{h}{c^\beta (\beta+1) (t/\xi_0)^\alpha}\right)
    \label{eq:Pht_alt1}
\end{equation}
where $f_\alpha$ is the L\'evy function.
Reintroducing the degree variable $k=h^{1/(1+\beta)}$ we find:
\begin{equation}
    P(k,t) = {k^\beta \over c^\beta(t/\xi_0)^\alpha}f_\alpha\left(
    \frac{k^{\beta+1}}{c^\beta (\beta+1) (t/\xi_0)^\alpha}\right),
    \label{eq:Pkt_alt1}
\end{equation}
that can be rewritten as:
\begin{equation}
    P(k,t) = {1 \over t^{\alpha /(1+\beta)}} \tilde f_{\alpha\beta}\left(
    \frac{k}{(t/\xi_0)^{\alpha/(1+\beta)}}\right) =
    {1 \over t^{\alpha /(1+\beta)}} \tilde f_{\alpha\beta} \left( \tilde k \right),
    \label{eq:Pkt_alt1_1}
\end{equation}
where $\tilde k = k/(c^{\beta} (\beta+1) (t/\xi_0)^{\alpha})^{1/(1+\beta)}$ and
$\tilde f_{\alpha\beta}(\tilde k)$ is an unknown function of $\tilde k$.

Equations (\ref{eq:Pkt_alt1_1}) and (\ref{eq:Pkt_alt1}) states that the peak of
the $P(k,t)$ distribution (i.e. the average degree) grows as
$t^{\alpha/(1+\beta)}$.

We can show these last findings in an alternative way. Prompted by the results
of the numerical simulations we suppose a scaling form of the $P(k,t)$ that
reads:
\begin{equation}
    P(k,t) \simeq {1 \over t^\gamma} P\left( {k^{1+\beta} \over t^\gamma}\right)
    \xrightarrow{h=k^{1+\beta}}
    P(h,t) \simeq {1 \over t^\gamma} P\left( {h \over t^\gamma}\right).
    \label{eq:scaling_alt1_1}
\end{equation}
If we now compute the space and time Fourier transform of Eq.
(\ref{eq:scaling_alt1_1}) calling $q$ the transformed variable of $h$ we find
\begin{equation}
    \begin{split}
    P(q,\omega) = \int\int e^{i\omega t + i q h} {1\over t^\gamma} P\left(
    {h\over t^\gamma} \right) dt dh =
    \int\int e^{it' + ih'} {1\over \omega}
    G\left( { h'/q \over (t'/\omega)^\gamma } \right) dt' dh'=
    {1\over\omega}\tilde G\left( {q\over \omega^\gamma} \right),
    \end{split}
    \label{eq:scaling_alt1_2}
\end{equation}
where we renamed $\omega t\to t'$ and $qh\to h'$.

By comparing the last term of the last line of Eq. (\ref{eq:scaling_alt1_2})
with Eq. (\ref{Pkt_alt1_1}) we find that $\gamma=\alpha$ so that we recover the
results shown in Eq. (\ref{eq:Pkt_alt1_1}).

The resulting scaling regime reads:
\begin{equation}
    P(h,t) =\frac 1 {c^\beta (\beta+1)(t/\xi_0)^{\alpha}} f_{\alpha\beta}\left(
    \frac{h}{c^\beta (\beta+1) (t/\xi_0)^\alpha}\right)
    \label{eq:Pkt_alt1_2}
\end{equation}
where $f_{\alpha\beta}$ is a non Gaussian scaling function.
Reintroducing the connectivity $k$ we get:
\begin{equation}
    P(k,t)= \frac {k^\beta} {c^\beta
    (t/\xi_0)^{\alpha}}f_\alpha\left(\frac{k^{\beta+1}}{c^\beta (\beta+1)
    (t/\xi_0)^\alpha}\right) =
    \left( {t\over \xi_0} \right)^{-\alpha \over \beta +1} \tilde
    f_{\alpha\beta}\left( {k \over \left( {t\over\xi_0} \right)^{\alpha/(\beta+1)}} \right),
    \label{eq:Pakt}
\end{equation}
where we grouped the $k/(t/\xi_0)^{\alpha/(\beta+1)}$ terms and where $\tilde
f_{\beta\alpha}(\tilde k)$ is an unknown , non-Gaussian scaling function of $k /
\left( {t\over\xi_0} \right)^{\alpha/(\beta+1)}$.

Moreover the overall time scaling of the average degree of an activity class
reads:
\begin{equation}
    k\sim c^{\beta/(1+\beta)}(a t)^{\alpha/(1+\beta)}
    \label{eq:kat}
\end{equation}
where we used the fact that $\xi_0$ is the inverse of the activity $a$.

\subsection{Asymptotic solution $1<\alpha<2$} 
\label{sub:Asymptotic solution_1a2}

First of all let us compute the functional form of the Fourier transformed
$P(k,t)$. As in the previous case, we find that the solution follows the:
\begin{equation}
    {1\over t^\gamma} P\left( {k^{1+\beta} - t \over t^\gamma} \right) =
    {1\over t^\gamma} P\left( {h - t \over t^\gamma} \right),
    \label{eq:scaling_1a2_1}
\end{equation}
where $h=k^{1+\beta}$. If we now follow the same procedure as in Equations
(\ref{eq:scaling_alt1_1}) and following, we find:
\begin{equation}
    \begin{split}
        \int{ e^{i\omega t + iqh} {1\over t^\gamma}
        P\left( {h - t \over t^\gamma}\right) dt dh } = \; [\epsilon = h-t] \; =
        \int{ e^{i\omega t + iq (\epsilon+t)} {1\over t^\gamma}
        P\left( {\epsilon \over t^\gamma} \right) dt dh } = \\
        [(\omega+q)t = t'; \, q\epsilon = q']\;\; =
        \int{ e^{it' + i\epsilon'} \left( {\omega+q\over t'} \right)^\gamma
        P\left({\epsilon'\over q}\left({\omega+q \over t}\right)^\gamma\right)
        {dt' \over \omega+q} {d\epsilon' \over q} } =
        {1\over \omega+q} g\left( {q\over (\omega+q)^\gamma} \right),\\
    \end{split}
    \label{eq:scaling_1a2_2}
\end{equation}
so that, calling $\tilde\omega=\omega+q$ we have $P(q,\tilde\omega) = {1\over
\tilde\omega} g\left( {q\over \tilde\omega^\gamma} \right)$.

Using Eq. (\ref{eq:intt_1a2}) we can rewrite Eq. (\ref{eq:Qkw_1}) as
\begin{equation}
    Q(k,\omega) = -\left( {c\over k} \right)^\beta {\partial Q \over \partial k}
    + {1\over 2} \left( {c\over k} \right)^\beta
    {\partial^2 Q \over \partial k^2} +
    i\omega\av{\xi} Q(k,\omega) + Q(k,\omega) +
    \omega^\alpha A_\alpha Q(k,\omega) +
    {\beta c^\beta \over k^{\beta+1}} Q(k,\omega) + \delta(k,0),
    \label{eq:Qkw_1a2_1}
\end{equation}
where $\av{\xi}$ is the first momentum of the inter event time distribution
(i.e. its mean value) and the second-last term comes from the expansion of the
$(c/(c+k-1))^\beta$ term.
We then introduce as in the previous case the $h=k^{\beta+1}$ variable, so that
$1/k^\beta \cdot \partial/\partial k \to \beta \partial/\partial h$ and
$1/k^\beta \cdot \partial^2/\partial k^2 \sim (1+\beta)^2 h^{\beta/(\beta+1)}
\cdot \partial^2/\partial h^2$. We then Fourier transform with respect to space
sending $h\to q$ getting:
\begin{equation}
    \begin{split}
    -i\omega\av{\xi} P(q,\omega) - A_\alpha \omega^\alpha P(q,\omega) =
    -i\beta c^\beta q P(q,\omega) + \\
    {1\over2} (1+\beta)^2 \int{e^{iqh} h^{\beta/(\beta+1)}
    {\partial^2P(h,\omega) \over \partial h^2} dh} +
    \beta c^\beta \int{h^{-1}e^{iqh} P(h,\omega) dh} + 1,
    \label{eq:Qkw_1a2_2}
    \end{split}
\end{equation}
where $P(h,\omega) = Q(h^{1/(1+\beta)}, \omega) = Q(k,\omega)$.
Now let us focus on the integral containing the second-derivative term. We can
rewrite it as:
\begin{equation}
    \begin{split}
    \int{e^{iqh} h^{\beta/(1+\beta)} {\partial^2\over\partial h^2}
    \int{e^{-i\omega t} P(h,t) dh dt}} =
    \iint{e^{iqh'} e^{-i\omega' t} (h'+\beta c^\beta t)^{\beta/(1+\beta)}
    {\partial^2\over\partial h'^2}{ P( (h'+\beta c^\beta t),t) dh' dt}},
    \end{split}
    \label{eq:Qkw_1a2_3}
\end{equation}
where we introduced $\omega' = \omega - \beta c^\beta q$ and $h'=h-\beta c^\beta
t$. Note that we can approximate $(h'+\beta c^\beta t) \simeq \beta c^\beta t$,
as we expect $h'\ll t$.
If we now integrate by parts Eq. (\ref{eq:Qkw_1a2_3}) and re-write
$P(h'+\beta c^\beta t, t)$ as $\int{e^{i\tilde\omega t  + i \tilde q h'}
P(\tilde q, \tilde\omega) d\tilde q d\tilde\omega}$ we get:
\begin{equation}
    \begin{split}
    -q^2 \iint{dh' dt e^{iqh'} e^{-i\omega' t} (\beta c^\beta t)^{\beta/(1+\beta)}
    \int{e^{i\tilde\omega t  + i \tilde q h'}
    P(\tilde q, \tilde\omega) d\tilde q d\tilde\omega}} =
    -q^2 \iint{dt d\tilde \omega e^{-i\omega' t + i\tilde\omega t}
    (\beta c^\beta t)^{\beta/(1+\beta)} P(q, \tilde\omega)},
    \end{split}
    \label{eq:Qkw_1a2_4}
\end{equation}
where we used the fact that the integrals in $dq$ and $d\tilde q$ give us a
$\delta(q,\tilde q)$.
Now we insert our prediction on the $P(q,\tilde\omega)$ of Eq. (\ref{eq:scaling_1a2_2}). In addiction we  call $\omega't=t'$ and, once the
substitution is done, $\tilde\omega/\omega' = y$ getting:
\begin{equation}
    -q^2 \iint{dy dt' e^{-it' + iyt'}
    (\beta c^\beta t'/\omega')^{\beta/(1+\beta)}
    \tilde\omega f\left( {q\over \tilde\omega^\gamma} \right)} =
    -{q^2 \over \omega'^{\beta/(1+\beta)}}
    H\left( {q \over \tilde\omega^\gamma} \right),
    \label{eq:Qkw_1a2_5}
\end{equation}
where $H(x)$ is an unknown scaling function.

Now, making the $\omega\to\omega'$ substitution and putting the result of Eq.
(\ref{eq:Qkw_1a2_5}) in Eq. (\ref{eq:Qkw_1a2_2}) we get:
\begin{equation}
    \begin{split}
        -i\omega'\av{\xi} f\left({q\over\omega^\gamma}\right) - A_\alpha (\omega'+\beta c^\beta q)^\alpha f\left({q\over\omega^\gamma}\right) + {1\over2} (1+\beta)^2 \left( {q \over \omega'^{ {1\over2}
    \beta/(1+\beta)}} \right)^2 H\left( {q \over \tilde\omega^\gamma} \right)
    = 1.
    \label{eq:Qkw_1a2_6}
    \end{split}
\end{equation}
Now we collect an $\omega'$ term from all the members to isolate the leading
order finding
\begin{equation}
    \begin{split}
        -i\av{\xi} f\left({q\over\omega^\gamma}\right) -
        A_\alpha \left(\omega'^{1-1/\alpha}
        +\beta c^\beta {q\over \omega'^{1/\alpha}}\right)^\alpha
        f\left({q\over\omega^\gamma}\right) +
        {1\over2} (1+\beta)^2 \left( {q \over \omega'^{ {1\over2}
        \left( {\beta\over (1+\beta) } + 1 \right)}} \right)^2
        H\left( {q \over \tilde\omega^\gamma} \right).
    \label{eq:Qkw_1a2_7}
    \end{split}
\end{equation}
Now if the term proportional to $A_\alpha$ is the leading one we find that the
exponent $\gamma=1/\alpha$, otherwise, if the term proportional to
$H(q/\omega'^\gamma)$ leads the evolution we find $\gamma =
(2\beta+1)/(2\beta+2)$.
The separation between the two cases is set at the point where the two exponents
are equal, e.g. the $A_\alpha$ term wins if
\begin{equation}
    \alpha < {2\beta+2 \over 2\beta+1}.
    \label{eq:alpha_beta}
\end{equation}
In the other case the term proportional to $H(q/\tilde\omega^\gamma)$ wins and
we recover the calculation of the previous work (see also Section
\ref{sub:asymptotic_solution_agt2}).

Given these results, to show the scaling form of the $P(k,t)$ distribution let
us recall the assumed scaling form of the $P(k,t)$ of Eq.
(\ref{eq:scaling_1a2_1}).  As already said the drift of the peak of the
distribution goes as $\av{k(t)}\propto t^{1/(1+\beta)}$, so that the variable
inside the distribution function reads:
\begin{equation}
    {k^{1+\beta}-vt \over t^{\gamma}}.
    \label{eq:scaling_general_1}
\end{equation}
If we rewrite $k^{\beta+1} = (k - t^{1/(1+\beta)} + t^{1/(1+\beta)})^{1+\beta}$,
we can introduce the variable $x = k-t^{1/(1+\beta)}\ll 1$ and expand Eq.
(\ref{eq:scaling_general_1}) in $x$:
\begin{equation}
    k^{1+\beta} \simeq t + (1+\beta)x t^{\beta/(1+\beta)} + \mathcal{O}(x^2).
    \label{eq:scaling_general_2}
\end{equation}
By substituting Eq. (\ref{eq:scaling_general_2}) in Eq.
(\ref{eq:scaling_general_1}) we find the distribution $P(k,t)$ to scale as:
\begin{equation}
    {1\over t^{\gamma - {\beta\over 1+\beta}}}
    P\left( {x \over t^{\gamma - {\beta\over 1+\beta}}}
    \right),
    \label{eq:scaling_general_3}
\end{equation}
where $x = k-t^{1/(1+\beta)}$.
Then in the Gaussian scaling $\gamma = {2\beta+1 \over 2(\beta+1)}$ so that the
scaling form reads:
\begin{equation}
    {1\over t^{1\over 2(1+\beta)}}P\left( {x \over t^{{1\over 2(1+\beta)}}}
    \right),
    \label{eq:scaling_gaussian}
\end{equation}
while in the other case $\gamma = {1 \over \alpha}$ so that the
scaling form reads:
\begin{equation}
    {1\over t^{ {1\over\alpha} - {\beta \over {1+\beta}}}}
    P\left( {x \over t^{ {1\over\alpha} - {\beta \over {1+\beta}} } }
    \right).
    \label{eq:scaling_levy}
\end{equation}

As a last remark, we stress that, regardless of the $\alpha$ and $\beta$
exponent, the peak of the distribution, i.e. the average value $\av{k(t)}$ grows
as
\begin{equation}
    \av{k(t)} \propto k^{1\over1+\beta}.
    \label{eq:avgk_1a2}
\end{equation}
This is in good agreement with numerical simulations and also allows for the
prediction of the degree distribution $\rho(k)$ as we will show in Section
\ref{sec:deg_dist}.

\subsection{Asymptotic solution $\alpha > 2$} 
\label{sub:asymptotic_solution_agt2}

Using Eq. (\ref{eq:intt_agt2}) we can rewrite Eq. (\ref{eq:Qkw_1}) as
\begin{equation}
    Q(k,\omega) = -\left( {c\over k} \right)^\beta {\partial Q \over \partial k}
    + {1\over 2} \left( {c\over k} \right)^\beta
    {\partial^2 Q \over \partial k^2} +
    i\omega\av{\xi} Q(k,\omega) + Q(k,\omega) +
    \omega^2 \av{\xi^2} Q(k,\omega) +
    {\beta c^\beta \over k^{\beta+1}} Q(k,\omega) + \delta(k,0),
    \label{eq:Qkw_agt2_1}
\end{equation}
where $\av{\xi^2}$ is the second moment of the $P(t)$ inter-event time
distribution.
The calculations in this case can be done in the direct space and we recover the
results of the previous work, in particular we find the $P(k,t)$ to scale as a
Gaussian with mean value growing as $(t/\xi_0)^{1/(1+\beta)}$ and variance
growing as $t^{1/[2(1+\beta)]}$:
\begin{equation}
    P(k,t) \simeq {1\over (t/\xi_0)^{1\over 2(1+\beta)}}
            \exp{\left[-A_{\beta}\frac{\left(k-C_\beta (t/\xi_0)^{{1\over
            1+\beta}}\right)^2}{(t/\xi_0)^{1/(1+\beta)}}\right]}.
    \label{eq:pkt_agt2_1}
\end{equation}
This results has the same functional form of the one found in Section
\ref{sub:Asymptotic solution_1a2} for the $\alpha>(2\beta+2)/(2\beta+1)$ case.



\section{Degree distribution $\rho(k)$} 
\label{sec:deg_dist}

The degree distribution can be evaluated recalling the scaling form accordingly
to the different phases of the system.
In particular we found:

\begin{equation}
    P(k,t)\simeq
        \begin{cases}
            {1 \over (t/\xi_0)^{\alpha\over1+\beta}}
            f_{\alpha\beta}\left( A'_{\alpha,\beta}
            {k \over (t/\xi_0)^{\alpha\over1+\beta}} \right)
            \,&\rm{if}\,\alpha<1,\\
            {1\over (t/\xi_0)^{ {1\over \alpha}-{1\over(1+\beta)} }}
            f_{\alpha\beta}\left( A'_{\alpha,\beta}{k-v(t/\xi_0)^{1/(1+\beta)}
            \over (t/\xi_0)^{ {1\over \alpha}-{1\over(1+\beta)} }}\right) &\,\rm{if}\,1<\alpha<{2\beta+2\over2\beta+1},\\
            {1\over (t/\xi_0)^{1\over 2(1+\beta)}}
            \exp{\left[-A_{\beta}\frac{\left(k-C_\beta (t/\xi_0)^{{1\over
            1+\beta}}\right)^2}{(t/\xi_0)^{1/(1+\beta)}}\right]}\,&\rm{if}\,\alpha>{2\beta+2\over2\beta+1},\\
        \end{cases}
    \label{eq:Pkt_total}
\end{equation}
so that we can evaluate, at fixed time $t$, the $\rho(k)$ distribution as:
\begin{equation}
    \rho(k) = \int_{\xi_{\rm min}}^{\xi_{\rm max}}{F(\xi_0) P(k,t) d\xi_0},
    \label{eq:rho_k_1}
\end{equation}
where $F(\xi)$ is the distribution of the lower-cut off $\xi$, i.e. the lower
cut-off of the inter-event time distribution for each agent.
If $F(\xi) = \delta(\xi-\xi_0)$ the $\rho(k)$ is trivially the $P(k,t)$.
If we instead let the $\xi$ of the system be distributed as $F(\xi)\propto
\xi^{\nu-1}$ we can evaluate the degree distribution $\rho(k)$.
Let us recall that such a distribution of the lower cut-off $\xi$ corresponds
to an activity distribution going as:
\begin{equation}
    F(a) \propto a^{-(\nu+1)},
    \label{eq:activity_distib}
\end{equation}
where we used the fact that $a\propto \xi^{-1}$.

For the $\alpha < 1 $ case we get:
\begin{equation}
    \begin{split}
    \rho(k) \propto \int{\xi^{\nu-1} \xi^{-\alpha/(1+\beta)}
    f_{\alpha\beta}\left( {k\over (t/\xi)^{\alpha/(1+\beta)}} \right) d\xi} = 
    \int{\xi^{\nu-1} \xi^{-\alpha/(1+\beta)}
    f'_{\alpha\beta}\left( {\xi k^{(1+\beta)/\alpha}} \right) d\xi} = \\
    [\xi' = \xi k^{(1+\beta)/\alpha}] =
    \int{k^{-1} \left({\xi' \over k^{(1+\beta)/\alpha}}\right)^{\nu-1}
    f'_{\alpha\beta}\left( {\xi'^{\alpha/(1+\beta)}} \right) {d\xi' \over
        k^{(1+\beta)/\alpha}}} = \mathcal{C} k^{-\left[{1+\beta \over \alpha}
        \nu + 1\right]},
    \end{split}
    \label{eq:rho_k_2}
\end{equation}
where $\mathcal{C}$ is a constant with respect to $k$.

For $\alpha>1$ we can use the scaling variable $x=k-t^{1/(1+\beta)}$ as a
$\delta(k-t^{1/(1+\beta)})$ so that we recover the result of the previous work:
\begin{equation}
    \rho(k) \propto k^{-\left[ (1+\beta)\nu + 1 \right]}.
    \label{eq:rho_k_3}
\end{equation}
